\documentclass[twocolumn,prb,notitlepage,superscriptaddress]{revtex4-2} 
\usepackage{amsmath,amsthm,amsfonts,graphicx,verbatim, xcolor,bm}
\usepackage{hyperref}
\usepackage[utf8]{inputenc}

\renewcommand{\vec}[1]{{\bm #1}}
\renewcommand{\Im}{{\rm Im}\,}

\newcommand{\ket}[1]{|#1\rangle}

\newcommand{\braket}[2]{\langle #1|#2\rangle}

\newcommand{\vect}[1]{\mathbf{#1}}

\begin{document}

\title{Geometric squeezing of rotating quantum gases into the lowest Landau level}
\author{Valentin Cr\'epel}
\affiliation{Center for Computational Quantum Physics, Flatiron Institute, New York, New York 10010, USA}
\author{Ruixiao Yao}
\affiliation{MIT-Harvard Center for Ultracold Atoms, Research Laboratory of Electronics, and Department of Physics, Massachusetts Institute of Technology, Cambridge, Massachusetts 02139, USA}
\author{Biswaroop Mukherjee}
\affiliation{MIT-Harvard Center for Ultracold Atoms, Research Laboratory of Electronics, and Department of Physics, Massachusetts Institute of Technology, Cambridge, Massachusetts 02139, USA}
\author{Richard J. Fletcher}
\affiliation{MIT-Harvard Center for Ultracold Atoms, Research Laboratory of Electronics, and Department of Physics, Massachusetts Institute of Technology, Cambridge, Massachusetts 02139, USA}
\author{Martin Zwierlein}
\affiliation{MIT-Harvard Center for Ultracold Atoms, Research Laboratory of Electronics, and Department of Physics, Massachusetts Institute of Technology, Cambridge, Massachusetts 02139, USA}

\begin{abstract}
The simulation of quantum Hall physics with rotating quantum gases is witnessing a revival due to recent experimental advances that enabled the observation of a Bose-Einstein condensate entirely contained in its lowest kinetic energy state, \textit{i.e.} the lowest Landau level. We theoretically describe this experimental result, and show that it can be interpreted as a squeezing of the geometric degree of freedom of the problem, the guiding center metric. This ``geometric squeezing'' offers an unprecedented experimental control over the quantum geometry in Landau-level analogues, and at the same time opens a realistic path towards achieving correlated quantum phases akin to quantum Hall states with neutral atoms. 
\end{abstract}

\maketitle

\section{Introduction} \label{sec:Introduction}

Quantum fluids and quantum gases under rotation exhibit a rich variety of phenomena, from Abrikosov vortex lattices~\cite{madison2000vortex,abo2001observation,Engels2003,zwierlein2005vortices} to quantum analogues of hydrodynamic instabilities and turbulence~\cite{sinha2001dynamic,sonin1987vortex,sonin2005ground,mukherjee2022crystallization}, with strong connections to other fields of physics such as rotating nuclei~\cite{Bohr1998,guery1999scissors}, neutron stars~\cite{pethick2015Bose,Neutronstars2019}, and electrons in high magnetic fields~\cite{wilkin2000condensation,cooper2001quantum}. At the core of this rich phenomenology is the interplay between two of the most fundamental properties of quantum matter: macroscopic quantum coherence -- manifest through superfluid behaviors -- and its coupling to a gauge field. Here, the gauge field is not dynamical but externally imposed by the rotation due to the identical mathematical structure of the Coriolis and Lorentz forces~\cite{Sivardiere_AnalogyInertialEM,fetter2009rotating}.

The observation of lattices of quantized vortices in Bose-Einstein condensates (BEC)~\cite{madison2000vortex} and later in strongly interacting Fermi gases~\cite{zwierlein2005vortices} provided a striking demonstration of superfluidity of quantum gases. Since these first demonstrations, one of the long-standing goal for these systems has been to increase the impact of the effective gauge field and reach the deeply degenerate regime, where quantum fluctuations are strong enough to coherently melt the vortex lattice~\cite{sinova2002quantum}. This occurs when all atoms live in their lowest kinetic energy manifold, which corresponds to the lowest Landau level (LLL) in the analogy with charged particles in magnetic fields, and when the total angular momentum of the atomic ensemble becomes comparable to the number of atoms. In this regime, the neutral atom analogue of integer and fractional quantum Hall states of electrons could potentially be realized~\cite{regnault2003quantum,regnault2004quantum,JolicoeurRegnault_Fermions,JolicoeurRegnault_Bosons}.

Moving in this direction, condensates with larger angular momentum were produced at ENS~\cite{SofteningLLL_Dalibard} and JILA~\cite{SofteningLLL_Cornell}, leading to vortex arrays containing hundreds of vortices and rotation near the lowest Landau level. Both of these experiments observed a softening of the vortex lattice, either through a qualitative change in appearance of the vortex lattice~\cite{SofteningLLL_Dalibard} or by direct measurement of the Tkachenko mode frequency~\cite{SofteningLLL_Cornell} that is related to the vortex lattice stiffness~\cite{baym2003Tkachenko}. This softening provided a promising precursor to the melting of the lattice induced at zero temperature by quantum fluctuations, and a deterministic route towards achieving the quantum Hall regime by reduction of the atomic density~\cite{sinova2003measuring}. 

To achieve such high angular momenta, the rotation frequency in these experiments was tuned as close as possible to the natural frequency of the underlying harmonic trapping potential~\cite{SofteningLLL_Cornell,SofteningLLL_Dalibard}. In fact, the physics of homogeneous electron gases is most directly realized when rotation and trapping frequency are equal~\cite{aftalion2009fast}. In this case indeed, the centrifugal force exactly compensates the harmonic confinement and the system is effectively in ``flat land'' where atoms are only subject to the effective gauge field imprinted by the Coriolis force. 

The lack of confinement in this regime stood as the main hurdle impeding further progress in the study of rapidly rotating quantum gases near the LLL regime~\cite{sinha2001dynamic,sinha2005two}. This difficulty shifted the focus to alternative ways to imprint an effective synthetic magnetic field on quantum gases –- using dressing by laser light~\cite{lin2009Bose,lin2009synthetic}, imprinting flux in optical lattices~\cite{struck2012tunable,aidelsburger2013realization,miyake2013realizing} or employing synthetic dimensions~\cite{celi2014synthetic,chalopin2020probing,fabre2022laughlin,crepel2020microscopic}. While these ideas have already led to elegant realizations of effective magnetic fields at the single-particle level~\cite{dalibard2011colloquium,goldman2014light,cooper2019topological}, their implementations in presence of interatomic interactions has so far been hindered by severe difficulties, such as heating in dressed or shaken optical lattices~\cite{reitter2017interaction} or all-to-all non-local couplings along synthetic dimensions that have been observed to energetically disfavor quantum Hall states~\cite{lkacki2016quantum}.

With the advent of single-atom-resolving microscopes for quantum gases~\cite{bakr2009quantum,sherson2010single,cheuk2015quantum,haller2015single,Parsons2015} and the ability to imprint arbitrary confining potentials~\cite{gaunt2013Bose,Mukherjee2017,zupancic2016ultra}, the original idea of employing rotation as the most direct analogue of the Lorentz force on charged particles is witnessing a revival. In a new experimental platform~\cite{fletcher2021geometric,mukherjee2022crystallization,yao2023observation}, our group has been able to directly image vortices in-situ, without time-of-flight expansion. Additionally, we have developed an alternative way of spinning up a quantum gas, entirely without introducing vortices, which we term \textit{geometric squeezing}. The present article aims to provide a theoretical account of this geometric squeezing and its consequences.

In essence, this new method harnesses the lack of confinement and the ensuing dynamical instability when spinning the gas at the trapping frequency, which originally prevented the observation of condensates in the LLL regime, to elongate the atomic cloud. This stretching simultaneously decreases the overall density of the system and increases its moment of inertia, and hence its angular momentum. As a result, condensates with typical angular momenta exceeding $1000h$ per particle and contained entirely in the LLL, in the form of a single Landau gauge wavefunction, are produced. This method provides an ideal starting point for the study of interactions within the LLL, both in the mean-field regime where the particle number largely exceed the number of vortices~\cite{mukherjee2022crystallization}; but also beyond mean-field as the number of atoms is reduced during geometric squeezing to become comparable to the number of flux lines, paving a new route for the realization of fractional quantum Hall states with neutral atoms~\cite{baranov2005fractional,cooper2013reaching,roncaglia2011rotating,crepel2019matrix,paredes2003fractional,leonard2023realization}. 

\section{Model and outline} \label{sec:Model}

We consider atoms of mass $m$ in a three-dimensional harmonic potential with natural frequencies $[\omega_x = \omega_\perp (1+\varepsilon), \omega_y = \omega_\perp (1-\varepsilon) , \omega_z]$ rotating around the vertical axis at angular velocity $\Omega$. For simplicity, we assume the axial dynamics completely frozen due to a strong vertical confinement $\omega_z \gg \omega_\perp$. The weak in-plane anisotropy, characterized by $\varepsilon$, imprints the rotation of the trap onto the atoms. We choose $\omega_\perp$, $\hbar \omega_\perp$ and $\ell_\perp = \sqrt{\hbar/(m\omega_\perp)}$ as respective units of frequency, energy and length. In the frame co-rotating with the trap, the single-particle dynamics of the system is governed by the Hamiltonian 
\begin{equation} \label{eq_Fermi3hamiltonian}
\mathcal{H} = \frac{1}{2} \left[ p_x^2 + p_y^2 + (1+\varepsilon) x^2 + (1-\varepsilon) y^2 \right] - \Omega L_z , 
\end{equation}
with $L_z = xp_y - yp_x$ the axial angular momentum~\cite{madison2000vortex}.

The last term in Eq.~\ref{eq_Fermi3hamiltonian} is responsible for the centrifugal and Coriolis fictitious forces. The latter has the same mathematical structure as the magnetic Lorentz force, which justifies the use of rotating gases to emulate the physics of charged particles in an magnetic field~\cite{Sivardiere_AnalogyInertialEM}. This is best seen by splitting $(- \Omega L_z)$ into two contributions, a deconfining potential $- \Omega^2 (x^2+y^2)/2$ and an effective vector potential $\vect{A} = \Omega [-y, x]$ equivalent to an applied magnetic field along the vertical direction:
\begin{equation} \label{eq_modelwithgauge}
\mathcal{H} = \frac{1}{2} \left[ (\vect{p}-\vect{A})^2 + (1 - \Omega^2 +\varepsilon) x^2 + (1 - \Omega^2 -\varepsilon) y^2 \right] \! ,
\end{equation}
which holds up to an overall constant.

The aim of the present article is to provide a theoretical description of the dynamical properties of Eq.~\ref{eq_Fermi3hamiltonian}, as experimentally observed in Refs.~\cite{fletcher2021geometric,mukherjee2022crystallization,yao2023observation}. For that purpose, we first review the single-particle properties of the model, starting with its dynamical instability near $\Omega = 1$ (Sec.~\ref{sec:SingleParticle}) and observe how the latter squeezes quantum states over time (Sec.~\ref{sec:squeezingsingleparticle}). This squeezing can be simply understood by the unitary evolution imposed by the rotating saddle potential, which physically implements a transformation from symmetric to Landau gauge in our system (Sec.~\ref{ssec:gaugetransformation}), and more intuitively explains the elongation of the quantum states and the reduction of the density that allows to reach the LLL. As a result of squeezing, overlaps between neighboring quantum states, which define the quantum geometry of the system~\cite{Haldane_FirstGeometryFQHE}, also change, as formally captured by a squeezing transformation of the guiding centers (Sec.~\ref{ssec:quantumgeometry}). This ``geometric squeezing'' provides a unique experimental control over the quantum geometry in Landau-level analogues. We finally connect this single-particle picture to a more realistic situation where interactions are accounted for using a hydrodynamic description of the superfluid (Sec.~\ref{sec:Hydrodynamics}) and full-fledged Gross-Pitaveskii numerical simulations of the condensate's dynamics (Sec.~\ref{sec:GPsection}).

\section{Classical solution} \label{sec:SingleParticle}

In this section, we study the dynamical instability of the Hamiltonian Eq.~\ref{eq_Fermi3hamiltonian} in more detail. We first locate the regime of instability, which is heralded by unbounded trajectories of the classical equations of motion (Sec.~\ref{ssec:classicalsolution}). We then interpret these unbounded solutions as a a guiding center drift following the isopotentials imprinted by the rotating saddle (Sec.~\ref{ssec:guidingcenterdrift}), and study the effects of this drift for a thermal phase-space distribution of particles (Sec.~\ref{ssec:phasespaceevolution}).

\subsection{Dynamical instability} \label{ssec:classicalsolution}

\begin{figure}
\centering
\includegraphics[width=\columnwidth]{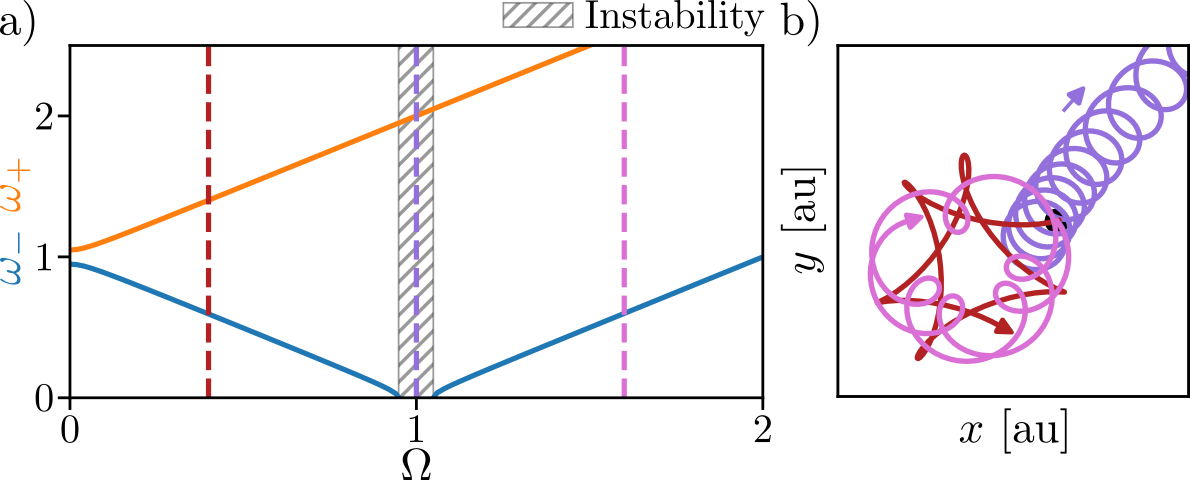}
\caption{a) Eigenfrequencies of the quadratic Hamiltonian Eq.~\ref{eq_Fermi3hamiltonian} for $\varepsilon=0.1$ as a function of the rotation frequency $\Omega$. The instability region $[\sqrt{1-\varepsilon},\sqrt{1+\varepsilon}]$ where $\omega_-$ becomes imaginary is hatched. b) Classical trajectories starting from the initial position marked as a black dot and $(p_x,p_y)|_{t=0} = (0,1)$. The color of the curve encodes the value of $\Omega$, which is also marked with a vertical dashed line in (a): red for $\Omega=0.4$, purple for $\Omega=1$, and pink for $\Omega=1.6$.}
\label{fig_classicalsolutions}
\end{figure}

To put Eq.~\ref{eq_Fermi3hamiltonian} in normal form and find its eigenmodes, we first decouple the position and momentum operators mixed by $L_z = xp_y - yp_x$. This is achieved by the following rotations admixing $(x, p_y)$ and $(y, p_x)$
\begin{equation} \label{eq_rotationcanonicaltransfo}
\begin{bmatrix} x' \\ p_y' \end{bmatrix} =  \begin{bmatrix} c & s \\ -s & c \end{bmatrix} \begin{bmatrix} x \\ p_y \end{bmatrix}, \quad \begin{bmatrix} y' \\ p_x' \end{bmatrix} = \begin{bmatrix} c & s \\ -s & c \end{bmatrix} \begin{bmatrix} y \\ p_x \end{bmatrix}, 
\end{equation}
with $c = \cos(\theta/2)$ and $s = \sin(\theta/2)$ and $\tan\theta = - 2 \Omega/\varepsilon$. Eq.~\ref{eq_rotationcanonicaltransfo} is a canonical transformation as it defines a new pair of conjugate variables $(x',p_x')$ and $(y', p_y')$. In terms of these new variables, the Hamiltonian can be split as $\mathcal{H} = \mathcal{H}_+ + \mathcal{H}_-$, where
\begin{equation} \label{eq_twodecoupledclassicaloscillator}
\mathcal{H}_+ = \frac{p_x'^2}{2m_+} + \frac{1}{2} k_+ x'^2 , \quad \mathcal{H}_- = \frac{p_y'^2}{2m_-} + \frac{1}{2} k_- y'^2 ,
\end{equation}
corresponds to harmonic oscillators with mass and coupling constant given by
\begin{align}
m_\pm^{-1} &= 1 \mp (\varepsilon / 2) \pm \sqrt{\Omega^2 + (\varepsilon / 2)^2} , \\
k_\pm &= 1 \pm (\varepsilon / 2) \pm \sqrt{\Omega^2 + (\varepsilon / 2)^2} .
\end{align}
While $k_+$ and $m_+$ are always positive, $k_-$ and $m_-$ respectively changes sign for $\Omega_- = \sqrt{1-\varepsilon}$ and $\Omega_+ = \sqrt{1+\varepsilon}$. When $\Omega \in [\Omega_-, \Omega_+]$, these coefficients have opposite sign and one of the system's eigen-frequencies 
\begin{equation}
\omega_\pm = \sqrt{k_\pm/m_\pm} = \left[ 1+\Omega^2 \pm \sqrt{\varepsilon^2+ 4\Omega^2} \right]^{1/2} ,
\end{equation}
becomes imaginary (see Fig.~\ref{fig_classicalsolutions}a), leading to a dynamical instability.

To illustrate this instability, let us integrate the classical equations of motion of the model. To that aim, we first obtain the time evolution operators of the two decoupled harmonic oscillators
\begin{equation} \label{eq_usingclassicalHO}
\begin{bmatrix} x'(t) \\ p_x'(t) \end{bmatrix} = U_+(t) \begin{bmatrix} x'(0) \\ p_x'(0) \end{bmatrix} , \quad \begin{bmatrix} y'(t) \\ p_y'(t) \end{bmatrix} = U_-(t) \begin{bmatrix} y'(0) \\ p_y'(0) \end{bmatrix}, 
\end{equation}
where standard calculations, repeated in App.~\ref{app_classical1dHO} for completeness, yield
\begin{equation} \label{eq_fulltimeevolution}
U_\pm (t) = \begin{bmatrix} \cos \omega_\pm t & \frac{\sin \omega_\pm t}{m_\pm \omega_\pm} \\ - \frac{k_\pm \sin \omega_\pm t}{\omega_\pm} & \cos \omega_\pm t \end{bmatrix} .
\end{equation}
Note that this result is valid for both real and imaginary frequencies $\omega_\pm$. The complete time evolution in terms of the original variables is then inferred from the rotations given in Eq.~\ref{eq_rotationcanonicaltransfo}. Some classical trajectories computed with these methods are displayed in Fig.~\ref{fig_classicalsolutions}b, where the black dot indicates the initial position from and $(p_x,p_y)|_{t=0} = (0,1)$. These trajectories clearly distinguish the stable regime with bounded trajectories (red and pink) from the dynamically unstable region characterized by unbounded trajectories (purple).

\subsection{Guiding center drift} \label{ssec:guidingcenterdrift}

A clear separation of scale can be observed when the rotation frequency matches the original trap frequency $\Omega=1$, where a slow drift along the first diagonal is superimposed to a much faster rotation of the particle (Fig.~\ref{fig_classicalsolutions}b). At this point, the centrifugal force exactly compensates the original confinement and the system is, in the rotating frame, equivalent to that of charged particles in a constant magnetic field subject to a saddle potential $\varepsilon (x^2 - y^2)/2$~\cite{fertig1987transmission}. The fast rotation corresponds to the cyclotron motion with period $2 \pi /\omega_+$, while the drift corresponds to the guiding center motion along the isopotential lines of the saddle~\cite{tong2016lectures}. Besides a stronger emphasis on the behaviors within the unstable regime, the classical solutions derived in this section and their interpretation in terms of cyclotron and guiding center motion are not new. They were, for instance, discussed in the context of anisotropic perturbations to the Foucault pendulum to explain the weak ellipticity of trajectories observed in some experiments~\cite{onnes1879nieuwe,airy1851vibration}.

\subsection{Generic phase space distributions} \label{ssec:phasespaceevolution}

Over lengthscales larger than the cyclotron radius, the effects of the fast and short-range cyclotron motion can be averaged out and the guiding center dynamics alone remains. Here, we use the classical time evolution obtained above to isolate and study the effects of guiding center drift on a classical -- or semi-classical -- phase-space distribution. 

We assume that the ensemble of particles, prepared using $\Omega = \varepsilon = 0$, can be described by a phase-space distribution $f_0(\vect{r}, \vect{p}) = f_0(E)$ that only depends on the local energy $E = (\vect{r}^2 + \vect{p}^2)/2$ with $\vect{r} = [x,y]^T$ and $\vect{p} = [p_x,p_y]^T$. Notably, this encompasses the Boltzmann, Fermi-Dirac and Bose-Einstein distributions, allowing us to describe classical, fermionic and bosonic ensembles at thermal equilibrium. However, our method is not limited to these cases and generically applies to all distributions that only depend on the classical local energy $E$ of the problem. 

The rotation $\Omega$ and ellipse $\varepsilon$ are turned on at $t=0$ to non-zero values, and the phase-space density $f_t(\vect{r}, \vect{p})$ at time $t>0$ can be obtained by following the classical trajectories of all particles in the ensemble. A particle found at phase-space point $(\vect{r},\vect{p})$ at time $t$ must have originated from the phase-space point $(\vect{r}(-t),\vect{p}(-t))$ at time $t=0$, so we have $f_t(\vect{r}, \vect{p}) = f_0(\vect{r}(-t), \vect{p}(-t)) = f_0(E_t)$, which only depends on the original energy $E_t = [\vect{r}(-t)^2 + \vect{p}(-t)^2]/2$ of the particle now found at $(\vect{r}, \vect{p})$. Because Eq.~\ref{eq_Fermi3hamiltonian} is quadratic, $E_t$ also is a quadratic form in the variables $(\vect{r}, \vect{p})$ that we formally write as 
\begin{equation} \label{eq_timeevolvedenergy}
E_{t} = \frac{1}{2} \begin{bmatrix} \vect{r} & \vect{p} \end{bmatrix} Q \begin{bmatrix} \vect{r} \\ \vect{p} \end{bmatrix} , \quad Q = \begin{bmatrix} Q_{rr} & Q_{rp} \\ Q_{pr} & Q_{pp} \end{bmatrix} .
\end{equation}
We provide the explicit form of $Q$ in App.~\ref{app_phasespacedensities} as determined from Eqs.~\ref{eq_rotationcanonicaltransfo} and~\ref{eq_fulltimeevolution}. 

\begin{figure}
\centering
\includegraphics[width=\columnwidth]{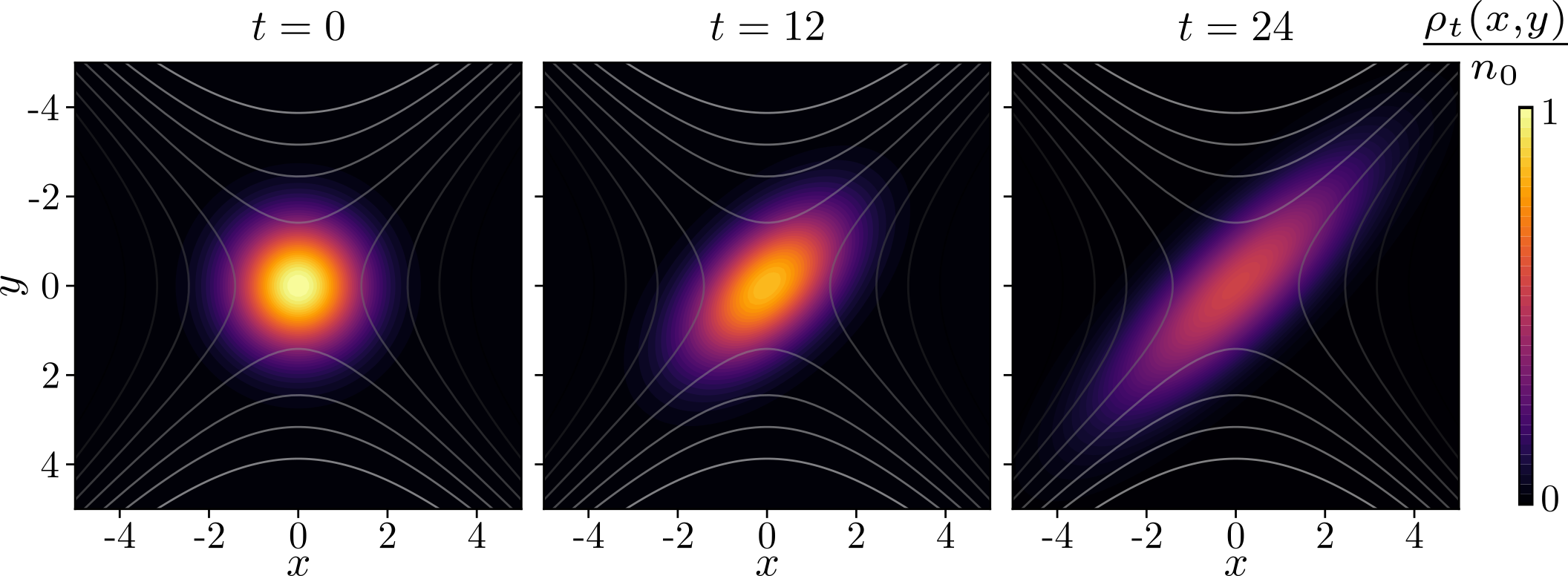}
\caption{Evolution of the real-space density for an ensemble originally described by a Boltzmann distribution with inverse temperature $(\beta = 1)$ as the rotation and anisotropy are switched on to $\Omega=1$ and $\varepsilon =0.1$ for $t>0$. As a guide to the eye allowing to visualize the guiding center drift, some isopotential lines of the rotating saddle potential $\varepsilon (x^2-y^2)$ are shown with solid lines using a gray scale, which goes from black (negative) to white (positive) values.
}
\label{fig_realspacedistributions}
\end{figure}

We are interested in the real-space density distribution 
\begin{equation}
\rho_t(\vect{r}) = \int {\rm d}^2 \vect{p} \, f_t(\vect{r}, \vect{p}) = \int {\rm d}^2 \vect{p} \, f_0 (E_t) , 
\end{equation}
which we compute using a linear transformation of the momenta consisting of a shift $\tilde{\vect{p}} = \vect{p} + Q_{pp}^{-1} Q_{pr} \vect{r}$ followed by a rotation and dilatation $\tilde{\vect{p}}_\theta = Q_{pp}^{1/2} \tilde{\vect{p}}$, with $Q_{pp}^{1/2}$ the square root of the symmetric matrix $Q_{pp}$. Relegating the lengthy but straightforward algebra to App.~\ref{app_phasespacedensities}, this procedure yields
\begin{eqnarray} \label{eq_densityrealspacedistribution}
\rho_t(\vect{r}) &=& \int {\rm d}^2 \tilde{\vect{p}} \, f_0 \left( \vect{r}^T Q_{pp}^{-1} \vect{r} + \tilde{\vect{p}}^T Q_{pp} \tilde{\vect{p}} \right) \\
&=& \frac{1}{\sqrt{|\det Q_{pp}|}} \int {\rm d}^2 \tilde{\vect{p}}_\theta \, f_0 \left( \vect{r}^T Q_{pp}^{-1} \vect{r} + \tilde{\vect{p}}_\theta^T \tilde{\vect{p}}_\theta \right) \notag \\ 
&=& \frac{\rho_0 \left( Q_{pp}^{-1/2} \vect{r} \right)}{\sqrt{|\det Q_{pp}|}} . \notag
\end{eqnarray}
It shows that the real-space density of the atomic ensemble keeps the same functional form in terms of a rotated and stretched coordinate, which results in elliptical equidensity lines. These ellipses are, up to an overall scale, entirely specified by the direction of their major axis, measured by its angle $\phi(t)$ from the $y$-axis, and the principal axis lengths $\lambda_\pm(t)$ given by the square roots of $Q_{pp}$'s eigenvalues. These parameters are derived in App.~\ref{app_phasespacedensities} using the explicit form of $Q_{pp}$, and are given by 
\begin{align} \label{eq_classicaldistribellipseparams}
\tan [2 \phi(t)] & = \frac{\Omega (c_+ \tau_+ - c_- \tau_-)}{ (\Omega_R + \Omega^2) \tau_+^2 + (\Omega_R - \Omega^2) \tau_-^2} , \\ 
\lambda_\pm^2 (t) & = 1 - \frac{\varepsilon^2 (\tau_+^2 - \tau_-^2)}{4 \Omega_R} \pm 
\frac{\varepsilon \Omega}{2\Omega_R} \left| \frac{c_-\tau_- - c_+ \tau_+}{\sin [2 \phi(t)] } \right| , \notag
\end{align}
where $c_\pm =\cos (\omega_\pm t)$, $\tau_\pm = \sin (\omega_\pm t) / \omega_\pm$, and $\Omega_R^2 = \Omega^2 + (\varepsilon/2)^2$.

In the dynamical instability and small anisotropy regime $(\Omega = 1, \varepsilon \ll 1)$, $\omega_-$ is imaginary such that the formula for $c_-$ and $\tau_-$ can be alternatively written in terms of $|\omega_-|$ with hyperbolic trigonometric functions. At long times, they therefore largely dominate in magnitude over $c_+$ and $\tau_+$, allowing to make analytical progress. In particular, we find that the tilt $\phi \simeq \arctan[-4/\varepsilon]/2 \simeq -\pi/4 +\varepsilon/8 \simeq -\pi/4$ brings the major axis of the distribution along the first diagonal. Similarly, the behavior of the major and minor axis length can be studied by writing $\lambda_\pm^2 = \kappa_\pm e^{2|\omega_-|t} + \alpha_\pm$ with $\alpha_\pm$ a function of time bounded by a constant, and $\kappa_\pm$ the coefficient corresponding to the instability. Expanding for long time, we get $\kappa_- = 0$ and $\kappa_+ = 1/(2\Omega_R) \simeq 1$. This shows that the minor axis remains a constant at long time while the major axis increases exponentially quickly at a rate $|\omega_-| = \varepsilon/2$. Altogether, the coefficients given in Eq.~\ref{eq_classicaldistribellipseparams} describe an exponential squeezing of the original rotation-symmetric cloud along the first diagonal, as captured by the long time behavior $\lambda_+ / \lambda_- \propto e^{\varepsilon t/2}$.

This is illustrated in Fig.~\ref{fig_realspacedistributions}, where we plot the density distribution $\rho_t$ at different times starting from a Boltzman distribution $\rho_0 (\vect{r})= n_0 e^{- \beta \vect{r}^2 / 2}$ of inverse temperature $\beta=1$, for which integration over momenta can be performed analytically. To make a closer connection with the guiding center drift discussed above, we overlay some isopotential lines of the rotating saddle, making clear that such drift is the fundamental reason behind the squeezing of the cloud.

\section{Squeezing quantum states} \label{sec:squeezingsingleparticle}

We now investigate the fate of a quantum state under the Hamiltonian Eq.~\ref{eq_Fermi3hamiltonian}, with a particular focus on the dynamically unstable regime identified above. Analogous to the classical case, single particle quantum states stretch out over time along the isopotential lines of the imposed rotating saddle (Sec.~\ref{ssec:squeezingsingleparticle}). In contrast to classical dynamics however, the zero point motion of the cyclotron harmonics imposes a minimum width to the density distribution even after an infinite evolution time. The quantum dynamics can be understood as physically effecting a transformation from the symmetric gauge to the Landau gauge (Sec.~\ref{ssec:gaugetransformation}), which arises from the evolution under the potential imprinted by the rotating saddle. We finally observe that the dynamics of our model can be formally described by a squeezing transformation of the guiding centers, which defines the quantum geometry in Landau-level analogues (Sec.~\ref{ssec:quantumgeometry}). As a result, the dynamical instability is a form of quantum ``geometric squeezing''.

\subsection{Explicit evolution of quantum states} \label{ssec:squeezingsingleparticle}

\subsubsection{Decoupling cyclotron and guiding center motion}

As in the classical case, we first decouple the normal modes of the Hamiltonian. While we could rely on the rotations used in Eq.~\ref{eq_rotationcanonicaltransfo} for that purpose, we notice that the decoupling can also be achieved by a simple gauge transformation. More precisely, we append the phase factor 
\begin{equation}
G = e^{i\kappa xy} , \quad \kappa = \varepsilon / (2\Omega) ,
\end{equation}
on all single particles states $\ket{\tilde\psi} = G\ket{\psi}$, which are now ruled by the Hamiltonian
\begin{align}
& \tilde{\mathcal{H}} = G \mathcal{H} G^\dagger \\ & = \frac{1}{2}[p_x^2 + p_y^2 +(1+\kappa^2)(x^2 +y^2)]  - \Omega L_z - \kappa (xp_y + yp_x) .  \notag
\end{align}
Introducing the cyclotron ($a_+$) and guiding center ($a_-$) bosonic operators, defined as 
\begin{equation} \label{eq_cyclotronandguidingcenter}
a_\pm = \frac{1}{2} \left[ \alpha (x \pm iy) + i \frac{p_x \pm ip_y}{\alpha} \right]  , \quad \alpha = (1+\kappa)^{1/4} ,
\end{equation}
the Hamiltonian separates into two independent parts $\tilde{\mathcal{H}} = \tilde{\mathcal{H}}_+ + \tilde{\mathcal{H}}_-$ that read 
\begin{equation} \label{eq_squeezingHamiltonian}
\tilde{\mathcal{H}}_\pm = \frac{\mu_\pm}{2} (2 a_\pm^\dagger a_\pm + 1) \pm \frac{\kappa}{2} ( a_\pm^2 + a_\pm^{\dagger \, 2} )  , 
\end{equation}
with $\mu_\pm = \sqrt{\omega_\pm^2 + \kappa^2}$, which describes the independent squeezing of the cyclotron and guiding center harmonic oscillators. 

\begin{figure*}
\includegraphics[width=\textwidth]{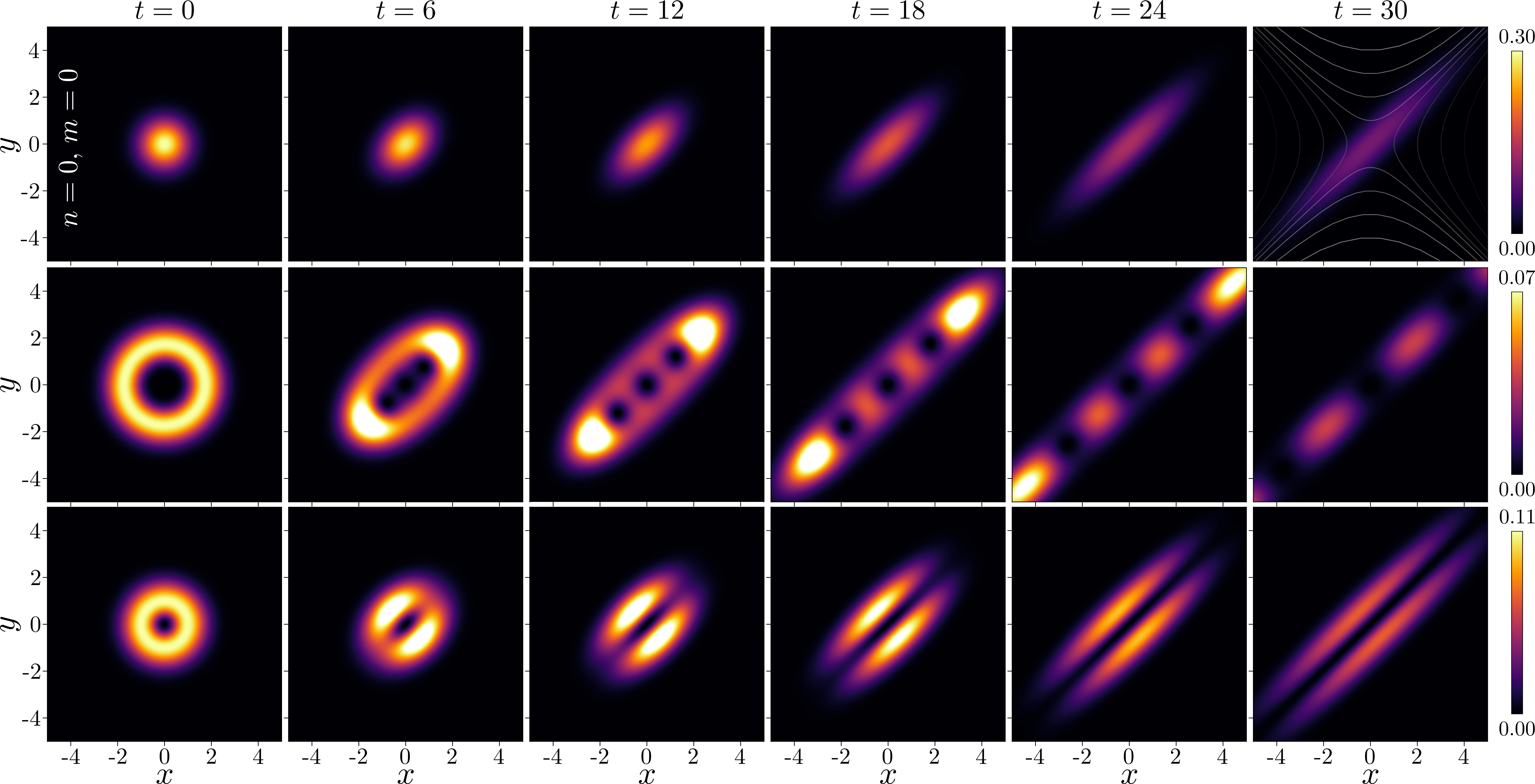}
\caption{
Density $|\braket{x,y}{n,m}_t|^2$ of the time evolved Fock states. After a small transient evolution, which is longer for higher cyclotron index $n$, the dynamics is well captured by isopotential flow of guiding centers. As a guide to the eye allowing to visualize the guiding center drift, some isopotential lines of the rotating saddle potential $\varepsilon (x^2-y^2)$ are shown on the topmost right panel with solid lines using a gray scale, which goes from black (negative) to white (positive) values.
}
\label{fig_FateOneBodyAll}
\end{figure*}

\subsubsection{Heisenberg evolution}

Before looking at the real-space representation of time-evolved wavefunctions, it is instructive to consider the evolution of the cyclotron and guiding center operators defined by
\begin{equation}
A_\pm(t) = \tilde{U}(t) a_\pm \tilde{U}^\dagger(t) , \quad \tilde{U}(t) = e^{-i t \tilde{\mathcal{H}}} .
\end{equation}
Using the Baker–Campbell–Hausdorff formula, we get 
\begin{align} \label{eq_heisenbergevolvedoperators}
A_\pm(t) & = f_\pm (t) a_\pm + g_\pm(t) a_\pm^\dagger , \\
f_\pm (t) & =  \cos \omega_\pm t + i \mu_\pm \frac{ \sin \omega_\pm t}{\omega_\pm} , \quad g_\pm(t) = \pm i \kappa \frac{ \sin \omega_\pm t}{\omega_\pm} . \notag
\end{align}
When compared to Eq.~\ref{eq_cyclotronandguidingcenter}, this explicit form of $A_+(t)$ suggests the definition of a novel time-dependent complex coordinate
\begin{equation} \label{eq_newholomorphic}
\xi(t) = \alpha [f_+(t) (x+iy) + g_+(t) (x-iy)] , 
\end{equation}
which drastically simplifies its expression
\begin{equation} \label{eq_cyclotrontimeevolved}
A_+(t) = \frac{1}{2} [\xi + 2ip_{\bar{\xi}}] ,
\end{equation}
where we have introduced $\bar\xi$ the complex conjugate of $\xi$, and $(p_\xi, p_{\bar\xi})$ the canonical momenta associated with $(\xi,\bar\xi)$; their explicit representation is provided in App.~\ref{app_Fockstates} for completeness. From Eq.~\ref{eq_cyclotrontimeevolved}, the physical interpretation of $\xi$ is clear: it defines the elliptic coordinate system most adapted to describe cyclotron orbits at any point in time. Note that we have defined $\xi$ using $A_+(t)$ to be sure that the method also applies in the regime of instability. Finally, we express the time-evolved guiding center operator using the new coordinate system
\begin{align} \label{eq_guidingtimeevolved}
A_-(t) & = \frac{1}{2} \left[ u(t) (\bar\xi + 2i p_{\xi}) - v(t) (\xi -2ip_{\bar\xi}) \right] , \\
u & = f_- f_+ - g_- g_+ , \quad v = f_- g_+^* - g_- f_+^* . \notag
\end{align}

\subsubsection{Quantum states}

Using this new system of coordinates, we can now efficiently determine the time evolution of arbitrary quantum states. For this, it is sufficient to find the evolution of a complete set of vectors at the initial time $(t=0)$. We consider two such sets: ($i$) the coherent states $\ket{\vec{\alpha}}_0$ satisfying $a_\pm \ket{\vec{\alpha}}_0 = \alpha_\pm \ket{\vec{\alpha}}_0$, and ($ii$) the Fock states $\ket{\vec{n}}_0$ diagonalizing the number operators $a_\pm^\dagger a_\pm \ket{\vec{n}}_0 = n_\pm \ket{\vec{n}}_0$. To obtain their time evolution, we rely on the fact that $\ket{\vec{\alpha}}_t = \tilde{U}(t) \ket{\vec{\alpha}}_0$ and $\ket{\vec{n}}_t = \tilde{U}(t) \ket{\vec{n}}_0$ can be determined, up to a global phase, as solutions of 
\begin{equation} \label{eq_timeevolvedeigenproblem}
A_\pm(t) \ket{\vec{\alpha}}_t = \alpha_\pm \ket{\vec{\alpha}}_t , \quad A_\pm^\dagger (t) A_\pm (t) \ket{\vec{n}}_t = n_\pm \ket{\vec{n}}_t .
\end{equation}

As a first example, let us derive the real-space representation of the time-evolved vacuum state defined by $A_\pm(t) \ket{0,0}_t = 0$. These relations provide two differential equations, which can be solved using a Gaussian ansatz, yielding
\begin{align} \label{eq_timeevolvedgroundstate}
\phi_t(\xi,\bar\xi) & \equiv \braket{x,y}{0,0}_t \\ &= \frac{1}{\sqrt{\pi |u|}} \exp\left[ \frac{\delta \xi^2 - |\xi|^2}{2} \right] , \quad \delta = \frac{v}{u} , \notag
\end{align}
where we have kept the time dependence of $\xi$, $u$ and $v$ implicit. The same approach can in fact be extended to any coherent state and leads to 
\begin{equation}
\braket{x,y}{\vec{\alpha}}_t =  \phi_t(\xi -\alpha_+,\bar\xi -\bar\alpha_+ - 2\alpha_-/u) e^{ -i\Im(\xi \bar\alpha_+) } .
\end{equation}
As in the most usual case, we observe that the coherent states are, up to a phase factor, shifted copies of the vacuum $\ket{0,0}_t$ obtained above. Finally, we can use the algebraic relations in Eq.~\ref{eq_timeevolvedeigenproblem} to express the time-evolved Fock states as
\begin{equation} \label{eq_algebraicdeffockstates}
\ket{\vec{n}}_t = \frac{1}{\sqrt{n_+! n_-!}} \left( A_+^\dagger \right)^{n_+} \left( A_-^\dagger \right)^{n_-} \ket{0,0}_t ,
\end{equation}
which provides, after lengthy calculations relegated to App.~\ref{app_Fockstates}, the explicit real-space representation of $\braket{x,y}{\vec{n}}_t$.

We finish this section by discussing more pictorially the time-evolution under $\mathcal{H}$. In Fig.~\ref{fig_FateOneBodyAll}, we show the density of the vacuum state -- which determines that of all other coherent states -- and of a few Fock states as a function of time for $\varepsilon=0.1$ and $\Omega=1$. The most striking feature is a drastic change in aspect ratio as a function of time. As is the case of classical distributions (Fig.~\ref{fig_realspacedistributions}). this can be understood as a result of particles flowing along the isopotential lines of the saddle potential, which are depicted on the upper right panel of Fig.~\ref{fig_FateOneBodyAll}. The main difference between the classical and quantum cases is the finite minor width that the quantum states still possess at long times. The latter is due to the zero point motion of the cyclotron operator, which sets the fundamental limit $\ell_B / \sqrt{2}$ on the width of the quantum states, with $\ell_B = \sqrt{\hbar / (2m\Omega)}$ the magnetic length.

\subsection{Effecting a gauge transformation} \label{ssec:gaugetransformation}

Focusing on a single Landau level, say the LLL, within which the kinetic energy is quenched, the time evolution under the saddle potential can be interpreted as performing a gauge transformation transforming symmetric gauge wavefunctions into Landau gauge ones, thus capturing the elongation of the states. Intuitively, this comes from the fact that the gauge transformation $U = e^{i\Omega xy}$ allows to transform the symmetric gauge used in Eq.~\ref{eq_modelwithgauge} into the Landau gauge, \textit{i.e.} $U \mathcal{H} U^\dagger$ has an effective gauge potential $\vect{A}' = 2 \Omega [0,x]$. Now, since the function $xy$ is identical to $(x^2 - y^2)/2$ after rotation of the axes by $\pi/4$, time-evolution under the saddle potential exactly reproduces the effect of the gauge transformation $U$. Note that this argument discards all kinetic contributions, which is justified within a given Landau level, where the kinetic energy is quenched by the stiff cyclotron harmonic oscillator. As a result, the time evolution under the rotating saddle potential physically implements a gauge transformation within the LLL. 

To see this more formally, let us project the original Hamiltonian (Eq.~\ref{eq_Fermi3hamiltonian}) for a rotation at the trap frequency $\Omega = 1$ onto its lowest energy level when $\varepsilon = 0$. This simply amounts to replacing $(x,y) \to (X,Y)$ with $(X,Y)$ the guiding center coordinates~\cite{tong2016lectures}, which are defined by 
\begin{equation}
X =  (x + p_y) / \sqrt{2}, \quad Y = (y - p_x)/ \sqrt{2} ,
\end{equation}
and match the $(p_y',y')$ defined in Eq.~\ref{eq_rotationcanonicaltransfo} for $\theta = -\pi/2$. After projection to the LLL, we thus get $\mathcal{P}_{\rm LLL} \mathcal{H} \mathcal{P}_{\rm LLL} = \varepsilon (X^2 - Y^2) / 2$ whose unitary evolution operator reads $U(t) = e^{-i \varepsilon t (X^2-Y^2)/2}$. Starting from the ground state in the symmetric gauge $\braket{x,y}{0,0}_0 = \exp[ -|z|^2/2]/\sqrt{\pi}$ with $z=x+iy$, we evolve it using $U(t)$ noting that $(X-iY)\ket{0,0}_0 =0$ and $[X,Y] = -i$ to find~\cite{fletcher2021geometric}
\begin{equation}
\braket{x,y}{0,0}_t = \frac{\exp \left[ - \frac{|z|^2 +i \tanh(\varepsilon t/2) z^2}{2} \right]}{\sqrt{\pi \cosh (\varepsilon t/2)}}  , 
\end{equation}
which is the same as Eq.~\ref{eq_timeevolvedgroundstate} for $\Omega = 1$ and $\kappa \ll 1$ if we assume the cyclotron motion unperturbed ($f_+=1$, $g_+=0$). For long times $\varepsilon t \gg 1$, it simply becomes 
\begin{equation}
\braket{x,y}{0,0}_t \simeq \frac{\exp[-i(x^2-y^2)/2 - (x-y)^2/2]}{\sqrt{\pi e^{\varepsilon t/2}} } ,
\end{equation}
that indeed describes a Landau gauge wavefunction of length $\sim e^{\varepsilon t/2}$ with its long axis rotated by $\pi/4$ compared to the original system of coordinates (see Fig.~\ref{fig_FateOneBodyAll}).

This simplified account of the dynamics provides two important insights. First, due to the quenched kinetic energy of the rapidly rotating quantum gas, the saddle potential effectively implements a gauge transformation through a unitary evolution. This evolution is coherent and does not introduce any heating nor turbulence in the form of disordered vortex nucleation. Second, the peak density of the cloud decreases exponentially with time, allowing to reach extremely dilute regimes that were proviously out of reach~\cite{fletcher2021geometric}, which opens a realistic path towards the long sought-after quantum Hall regime of rotating quantum gases.

\subsection{Relation to quantum geometry} \label{ssec:quantumgeometry}

\begin{figure}
\centering
\includegraphics[width=\columnwidth]{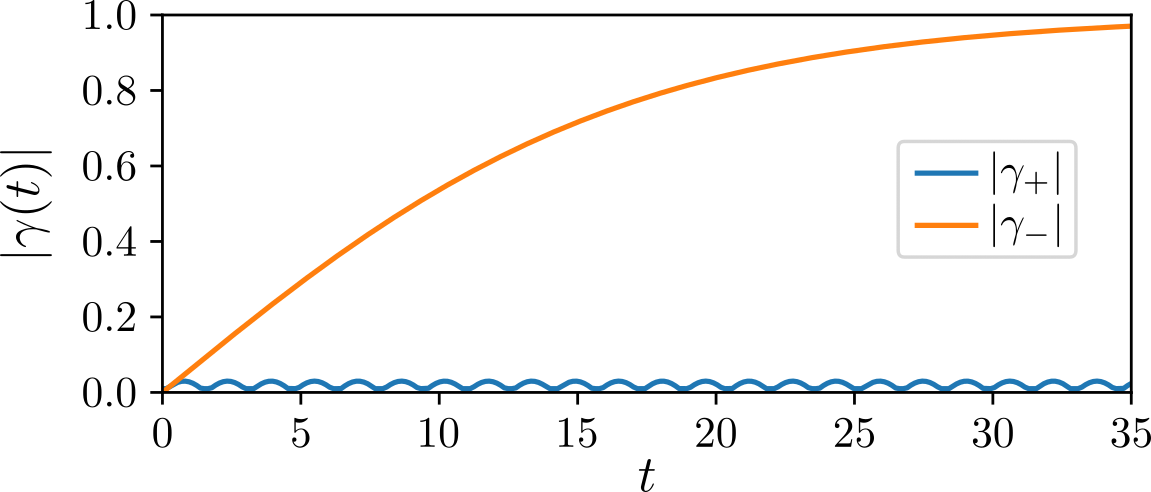}
\caption{Amplitude of the squeezing parameters for the cyclotron $\gamma_+$ and guiding center $\gamma_-$ operators when rotation and anisotropy are switched on at $t=0$ to $\Omega = 1$ and $\varepsilon =0.1$. Except for a weak breathing at frequency $\omega_+$ due to the cyclotron motion, most of the observable features, \textit{e.g.} in Fig.~\ref{fig_FateOneBodyAll}, are attributable to the guiding center dynamics.}
\label{fig_squeezingparameters}
\end{figure}

After a small transient evolution, the full quantum evolution seems to be explained by guiding center drift along the saddle, with almost no appreciable effects attributable to the cyclotron degree of freedom. This might, at first sight, seem odd since both cyclotron and guiding center motion are ruled by similar squeezing Hamiltonians (Eq.~\ref{eq_squeezingHamiltonian}). Looking more closely at the squeezing parameter 
\begin{equation}
\gamma_\pm = g_\pm / f_\pm , 
\end{equation}
of the two operators, plotted as a function of time in Fig.~\ref{fig_squeezingparameters}, reveals that the cyclotron operator is very stiff $\omega_+ \gg \kappa$ and only slightly breathes at its natural frequency without experiencing much squeezing. On the contrary, $\omega_-$ is of the order of or even smaller than the squeezing amplitude $\kappa$, leading to a much larger squeezing parameter $\gamma_- \gg \gamma_+$ that reaches values close to unity for long times. Most of the observable features in the time evolution of the system are thus directly attributable to the guiding center dynamics.

This is a generic feature of well-separated Landau levels, which possess a stiff cyclotron mode and a much softer guiding center degree of freedom. In fact, it is known that, in presence of an external perturbation, only the soft guiding center mode will adjust to accommodate the perturbation. It will do so in a purely geometric manner, by changing the local shape of the guiding center coherent states. This geometrical response of a quantum Hall system was first spotlighted by Haldane, which described the quantum geometric degree of freedom of the system in terms of a \emph{guiding center metric} $g$~\cite{Haldane_FirstGeometryFQHE}, which in our notation reads~\cite{Haldane_ModelAnisotropic}
\begin{equation}
g = \frac{1}{1-|\gamma_-|^2} \begin{bmatrix} |1+\gamma_-|^2 & i (\gamma_- - \gamma_-^*) \\ i (\gamma_- - \gamma_-^*) & |1-\gamma_-|^2 \end{bmatrix} . 
\end{equation}

In the context of quantum Hall systems, geometric responses of the guiding-center metric have already been observed for an applied in-plane magnetic field~\cite{Papic_AnisotropiPseudoPotential} or an anisotropic band mass tensor~\cite{HaldanePapic_BandMassAnisotropy}. In another context, a spontaneous symmetry breaking of a fractional quantum Hall system toward a nematic phase, \textit{i.e.} a transition from $\gamma=0$ to $|\gamma|\simeq 1$, was predicted as a way to minimize the interaction energy~\cite{RegnaultFQHENematic}. Beyond these geometric responses due to the guiding center degree of freedom, the relation between quantum geometry and guiding center metric was observed to be more profound~\cite{Haldane_FirstGeometryFQHE}. Indeed, in condensed-matter systems, bands with non-zero Chern number that analytically reproduce the physics of Landau levels were proved to only have one degree of freedom: their quantum metric, which also characterizes the guiding center metric of the Landau level they map onto~\cite{estienne2023ideal}.

Here, we have shown how the dynamical instability of rotating quantum gases could be used to modify the quantum geometry of the system by squeezing of the guiding centers. This ``geometric squeezing'' experimentally pioneered in Ref.~\cite{fletcher2021geometric} results in an unprecedented control over the quantum geometry of the system, and offers a new way to reach the lowest Landau level.

\section{Hydrodynamic description of geometric squeezing} \label{sec:Hydrodynamics}

The previous analysis completely neglected the effects of interatomic interactions. However, for the experimental conditions of Ref.~\cite{fletcher2021geometric}, the typical interaction energy at initial time is much larger than the cyclotron gap and the density profile of the condensate at small time is mostly governed by interaction effects. We now develop a hydrodynamic description of the condensate in the presence of these interactions and solve for its dynamics in absence of quantum pressure.

\subsection{Hydrodynamic equations}

The dynamics of a condensate within a rotating trap $U(r)$ is amenable to a hydrodynamic description, obtained by rewriting the Gross-Pitaevskii (GP) equation for the wavefunction $\Psi = \sqrt{\rho}\, e^{i S}$ as equivalent hydrodynamic equations for the density $\rho = |\psi|^2$ and the superfluid phase $S$. As in Sec.~\ref{sec:Model}, we work in the rotating frame, where the GP equation reads~\cite{Review_BoseGasInTraps}
\begin{equation}
i\hbar \partial_t \Psi = \left[\frac{-\hbar^2 \nabla^2}{2m}+U(\vect{r})-\Omega {L}_z \right] \Psi + g |\Psi|^2 \Psi , 
\end{equation}
with $g$ the interaction coefficient, and substitute in $\psi = \sqrt{\rho}\, e^{i S}$. Taking the imaginary part yields
\begin{eqnarray}
\frac{\partial \rho}{\partial t} &=& -\nabla\cdot\left(\rho (\vect{v}- \vect{\Omega}\times\vect{r})\right), \quad \vect{v} = (\hbar \nabla S) / m ,
\end{eqnarray}
which is the continuity equation in the rotating frame; while taking the real part yields 
\begin{eqnarray}
-\hbar \frac{\partial S}{\partial t} =&-& \frac{\hbar^2}{2m} \frac{1}{\sqrt{\rho}} \nabla^2\sqrt{\rho} + \frac{\hbar^2}{2m}(\nabla S)^2 + U(\vec{r})\nonumber\\
 &-& \hbar\vect{\Omega}\cdot(\vect{r}\times(\nabla S)) + g \rho.
\end{eqnarray}
As noted in~\cite{sinha2001dynamic}, for a condensate in a harmonic trap these equations can be solved analytically via a quadratic ansatz for the superfluid wavefunction. It is necessary however to neglect the quantum pressure term $\sim\nabla\sqrt{\rho}$, which is valid when it is dominated by the mean-field term $g\rho$. The smallest lengthscale across which the superfluid wavefunction may vary is set by the magnetic length, corresponding to the spatial extent of cyclotron orbits in the lowest Landau level. This means that the relative importance of the quantum pressure and mean-field terms is simply given by the ratio of the chemical potential to the Landau level spacing, and loosely corresponds to the number of Landau levels admixed into the superfluid wavefunction. We therefore expect a classical hydrodynamic description to be valid when this number is much larger than unity.

\subsection{Analytic solution}

Following~\cite{sinha2001dynamic}, to which we refer for more details, we consider an anisotropic harmonic trap of the form
\begin{equation}
    U(\vect{r})=\frac{1}{2}m\omega_\perp^2\left(\left[1+\varepsilon\right]x^2+\left[1-\varepsilon\right]y^2+\left(\frac{\omega_z}{\omega_\perp}\right)^2 z^2\right) ,
\end{equation}
and make a quadratic ansatz for the density and phase,
\begin{eqnarray}
    \rho(\vect{r},t)=\rho_0(t)+\frac{m\omega_\perp^2}{g}\sum\limits_{i,j=1}^3 x_i A_{ij}(t) x_j,\nonumber\\
    S(\vect{r},t)=S_0(t)+m\omega_\perp\sum\limits_{i,j=1}^3 x_i B_{ij}(t) x_j.
\end{eqnarray}
The time-evolution of the condensate wavefunction is thus encoded in the matrices $A$ and $B$, which evolve according to~\cite{sinha2001dynamic}
\begin{eqnarray}
    \frac{1}{\omega_\perp}\frac{\textrm{d}A}{\textrm{d}t}&=&-2A\,\textrm{Tr}B-2(AB+BA)+\frac{\Omega(t)}{\omega_\perp}(RA-AR)\nonumber \\
    \frac{1}{\omega_\perp}\frac{\textrm{d}B}{\textrm{d}t}&=&-2B^2-W-A+\frac{\Omega(t)}{\omega_\perp}(RB-BR),
\end{eqnarray}
where the matrices $W$ and $R$ are defined as
\begin{eqnarray}
W=\frac{1}{2}\begin{pmatrix}
1+\varepsilon & 0 & 0\\
0 & 1-\varepsilon & 0\\
0 & 0 & (\omega_z/\omega_\perp)^2
\end{pmatrix},
\end{eqnarray}
and 
\begin{eqnarray}
R=\begin{pmatrix}
0 & 1 & 0\\
-1 & 0 & 0\\
0 & 0 & 0
\end{pmatrix}.
\end{eqnarray}

\begin{figure}
\centering
\includegraphics[width=\columnwidth]{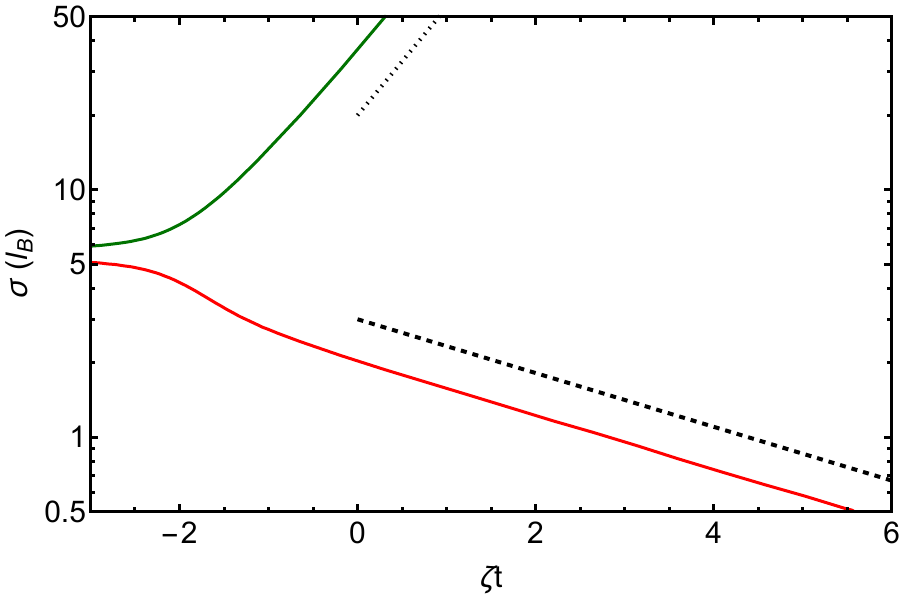}
\caption{The evolution in the major and minor radii of a condensate held in a rotating harmonic trap. While the long axis grows exponentially with a rate corresponding to the squeezing of guiding center coordinates (dotted line), the minor width falls more slowly (dashed line). This is due to the falling density leading to smaller Landau level occupation, and hence a smaller size of cyclotron orbits (see text).}
\label{fig_hydro}
\end{figure}

This formalism allows straightforward calculation of the condensate dynamics under arbitrary variation in the trap rotation frequency. As an example, we show in Fig.~\ref{fig_hydro} the evolution in the condensate $e^{-1/2}$ radii along its major and minor axes, $\sigma_+$ and $\sigma_-$, for the experimental parameters of~\cite{fletcher2021geometric}. An initially equilibrium cloud was prepared in an anisotropic trap with $\varepsilon=0.125$ and $\omega_\perp=2\pi\times88.6~$Hz, whose rotation rate was smoothly increased from $\Omega=0$ to $\Omega=\omega_\perp$ and held for a variable time $t$. 

While the long cloud axis grows exponentially $\sim\exp(\zeta t)$ where $\zeta=\varepsilon\omega_\perp/2$, the short axis falls more slowly. 
This is because the cloud size contains contributions from both the guiding centers, which are squeezed at a rate $\zeta$, and from the cyclotron orbits, whose size depends upon the number of occupied Landau levels $N_\textrm{LL}\equiv\mu/(2\hbar\omega_\perp)$. 
The squeezing of guiding centers means that for most of the experiment $\sigma_-$ is generally dominated by cyclotron motion and its evolution is captured by a simple scaling model. 
The chemical potential is proportional to the atomic density $\sim 1/(\sigma_+\sigma_-\sigma_z)$, where $\sigma_z$ is the axial extent of the condensate. The major width always increases as $\sigma_+\propto \exp(\zeta t)$, and the short axis size $\sigma_{-,z}\propto\sqrt{\mu}$ when $N_\textrm{LL}\gg1$. We therefore predict a time-dependence $\sigma_-\propto \exp(-\zeta t/4)$, which is shown by the dashed line in Fig.~\ref{fig_hydro}. 
 
The falling chemical potential $\mu\propto \exp(-\zeta t/2)$ guarantees that eventually $\mu<2\hbar\omega_\perp$ and the condensate enters the LLL. In the experiment of~\cite{fletcher2021geometric} and the GP simulation of Fig.~\ref{fig_gs}, we find that $\sigma_-$ saturates at the zero-point cyclotron orbit size imposed by Heisenberg uncertainty. 
However, since the hydrodynamic model neglects quantum pressure, it instead predicts that $\sigma_-\rightarrow 0$.

\section{Gross-Pitaevskii simulations} \label{sec:GPsection}

A clear picture of how geometric squeezing enables us to reach the lowest Landau level emerges from the non-interacting treatment presented above: due to the quenched kinetic energy of the rapidly rotating quantum gas, the time evolution operator is mainly due to the anisotropic saddle potential that physically implements a symmetric-to-Landau gauge transformation (Sec.~\ref{ssec:gaugetransformation}). This evolution squeezes the atomic cloud, which simultaneously decreases the peak density exponentially with time and increases its moment of inertia. This reduces the effects of interactions (Sec.~\ref{sec:Hydrodynamics}), forming the ideal conditions to reach the LLL. 

We perform full Gross-Pitaveskii numerical simulations of the dynamics of the condensate, in the presence of interactions. The formalism and numerical details are gathered in Sec.~\ref{ssec:GPdetails}, while the results of the simulations are discussed in Sec.~\ref{ssec:GPresults}

\subsection{Formalism} \label{ssec:GPdetails}

We now consider a BEC containing $N$ atoms in the rotating anisotropic trap of Eq.~\ref{eq_Fermi3hamiltonian} described by the macroscopically occupied mode $\Psi$. Within the purely two-dimensional settings introduced in Sec.~\ref{sec:Model}, the GP equation takes the form~\cite{Review_BoseGasInTraps}:
\begin{equation} \label{eq_GPequation}
i \partial_t \Psi (\vec{r}, t) = \left[ \mathcal{H} + g_{\rm 2d} |\Psi (\vec{r},t)|^2 \right] \Psi (\vec{r}, t) , 
\end{equation}
where $g_{\rm 2d} = N a_s \sqrt{8\pi \omega_z}$ is the effective effective two-dimensional interaction parameter, with $a_s$ the scattering length of the atoms~\cite{Shlyapnikov_QuasiTwoDimScatteringLength}.

To integrate the time-dependent GP equation, we use a time splitting pseudospectral method (TSSP)~\cite{ruprecht1995time}. This method is most often used in absence of rotation, and harnesses the fact that the kinetic term $\propto \vec{p}^2$ is easily evolved in Fourier space while the interaction and potential terms are most easily treated in real space~\cite{dion2003spectral}. The TSSP uses a Trotter decomposition of the time-evolution operator, and the infinitesimal time evolution operator at each timestep is split to separate the kinetic term from the rest of the Hamiltonian. This enables us to Fourier transform $\Psi$ before evolving it with the kinetic terms, leading to local time evolution operators that can be efficiently implemented numerically~\cite{bao2013mathematical}. 

In the rotating frame, the angular momentum operator $L_z = xp_y - yp_x$ explicitly mixes the coordinate and momentum operators, and the split-step method should be modified for best performance. More precisely, we first implement a one-dimensional Fourier transform to bring $\Psi(x,y,t)$ into $\Psi(k_x,y,t)$ and evolve it under the $\Omega y p_x$ part of the Hamiltonian. Fourier transforming both axes, we arrive at $\Psi(x,k_y,t)$, which is evolved under $\Omega x p_y$ piece of the Hamiltonian. We alternate the order of these evolutions, implementing $yp_x$ first for even timesteps and $xp_y$ first in odd ones, which further reduces systematic computational errors.

\subsection{Evolution under geometric squeezing} \label{ssec:GPresults}

\begin{figure}
\centering
\includegraphics[width=\columnwidth]{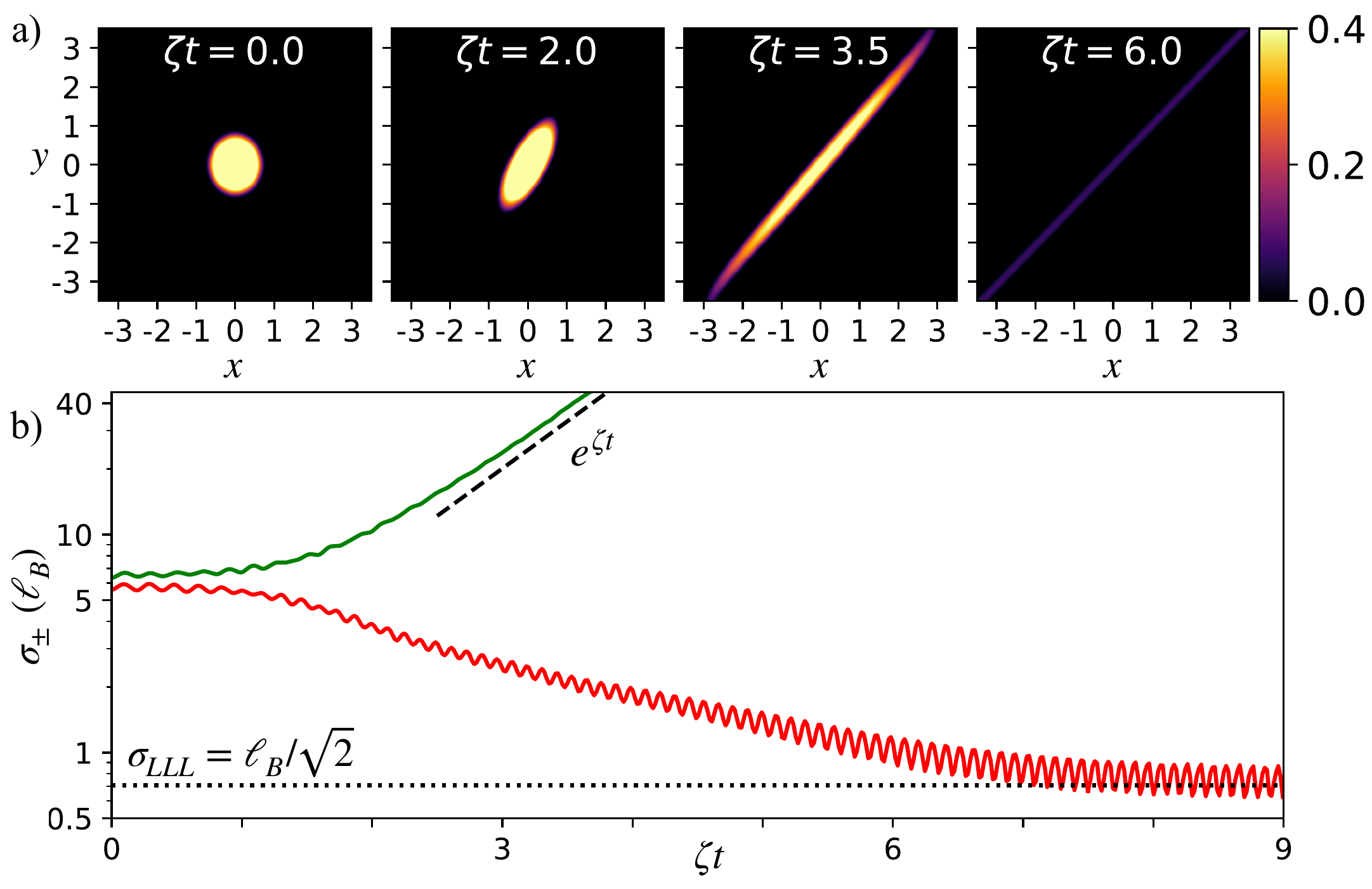}
\caption{a) Density of the condensate wave-function obtained by numerical integration of the GP equation for $\varepsilon=0.125$ and a ramp from $\Omega(t=0)=0$ to $\Omega (t = \infty)=1$ that matches the experimental sequence of Ref.~\cite{fletcher2021geometric}. b) Major ($\sigma_{+}$, green) and minor ($\sigma_{-}$, red) typical width of the condensate wavefunction as a function of time. Time is normalized by the squeezing rate, $\zeta t$ with $\zeta = \varepsilon/2$, such that the exponential increase of the major width at long time becomes universal (affine with a slope of one in the semi-logarithmic used here). Finally, the dotted horizontal line shows indicates the zero-point width $\ell_B/\sqrt{2}$ of a Landau gauge orbital, to which $\sigma_{-}$ saturates.}
\label{fig_gs}
\end{figure}

An example of GP simulation is shown in Fig.~\ref{fig_gs}. We initially prepare a non-rotating and weakly interacting BEC at equilibrium in the anisotropic trap of Eq.~\ref{eq_Fermi3hamiltonian}, 
where $\varepsilon=0.125$ resembles the experimental settings. We then ramp up the rotation frequency from $\Omega(t=0)=0$ following a sequence similar to the experimental procedure of Ref.~\cite{fletcher2021geometric}. This gradually brings the BEC gradually to rotate at the trap frequency $\Omega = 1$. 

The density profiles displayed in Fig.~\ref{fig_gs}a clearly demonstrate that, as in the non-interacting case, the BEC elongates along the first diagonal in the rotating frame as a result of the guiding center flow following the isopotential of the rotating saddle. To be more quantitative, we plot in Fig.~\ref{fig_gs}b the $e^{-1/2}$ radii of the condensate wavefunction along its major ($\sigma_{+}$) and minor ($\sigma_{-}$) axes as a function of time. 
Consistent with the squeezing of the guiding centers, $\sigma_{+}$ grows exponentially at a rate set by $\varepsilon/2$. The minor axis is also exponentially reduced at early times, but eventually saturates around the value $\ell_B/\sqrt{2}$, the width of a Landau gauge wavefunctions. This clearly signals that, as the peak density of the BEC decreases, the chemical potential of the cloud becomes comparable to the cyclotron gap and all of the atom are eventually contained into the LLL, in the form of a macroscopically occupied Landau gauge wavefunction of minor width $\ell_B/\sqrt{2}$. We finally note that oscillations in $\sigma_-$ at frequency $2\omega_+$, clearly visible in Fig.~\ref{fig_gs}b, are also present in the non-interacting solutions of Sec.~\ref{sec:squeezingsingleparticle} and are due to the breathing of the cyclotron degree of freedom, explicit in the oscillatory nature of $\gamma_-(t)$ in Fig.~\ref{fig_squeezingparameters}. 

\begin{figure*}
\centering
\includegraphics[width=\textwidth]{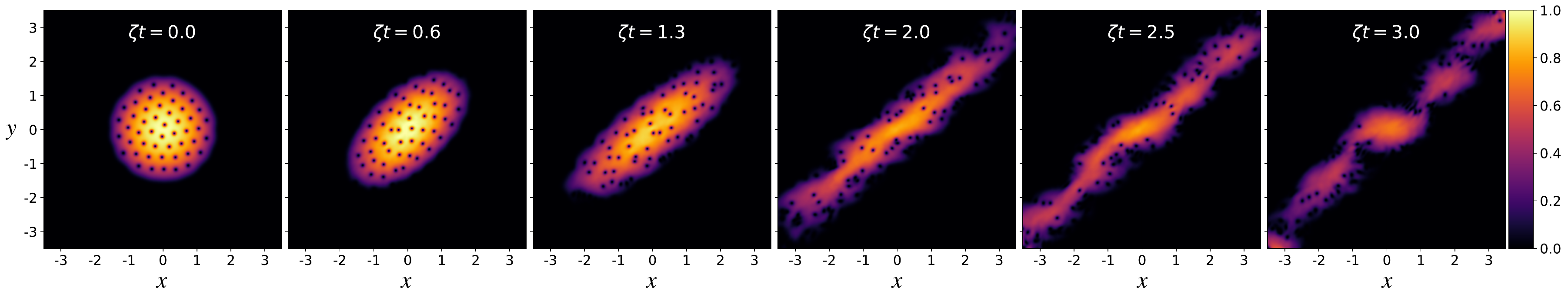}
\caption{The density profile of the vortex lattice under geometric squeezing. Squeezing is manifested in the cloud profile and the vortex distribution.}
\label{fig_vlat}
\end{figure*}

As another example, we show in Fig.~\ref{fig_vlat} the evolution of a vortex lattice under the same evolution. At initial time, the vortex lattice is equilibrated with in a 2D harmonic trap rotating at $\Omega = 0.8$. The anisotropy of the trap is switched on at $t=0$ and keeps rotating at $\Omega$. In the rotating frame where the saddle potential is static, the vortex lattice elongates along the isopotential diagonal, with the vortices moving along the same direction. The flow of vortices at short times indicate that only the guiding center motion evolves and gets squeezed, leaving the cyclotron motion unchanged. For longer squeezing times, the inter-vortex distance becomes comparable to the minor axis of the cloud and their dynamics strongly couple to the overall squeezing of the atomic cloud, leading to a complex pattern formation.

\section{Conclusion} \label{sec:Conclusion}

In this article, the ``geometric squeezing'' experimentally pioneered in Ref.~\cite{fletcher2021geometric} to bring rotating quantum gases into the lowest Landau level has been theoretically studied using a combination of analytical arguments and numerical simulations. At its core, this method uses novel experimental improvements to harness the dynamical instability that originally prevented the observation of quantum gases in the lowest Landau level. This instability is due to the unbounded trajectories of the guiding center flowing along the isopotential lines of the saddle potential imposed by the rotating anisotropic trap. This evolution elongates the atomic ensemble, thereby simultaneously decreasing its peak density and increasing its moment of inertia. This naturally lead to a dilute gas with low interaction energy and high angular momentum per particle, which is entirely contained within the lowest Landau level in the form of a macroscopically occupied Landau gauge wavefunction.

\section*{Acknowledgement}

This work was supported by the NSF through
the Center for Ultracold Atoms and Grant PHY-2012110, and by ARO.
M.Z. acknowledges funding from the Vannevar Bush Faculty Fellowship (ONR N00014-19-1-2631). R.J.F. acknowledges funding from the AFOSR Young Investigator
Program (FA9550-22-1-0066).
The Flatiron Institute is a division of the Simons Foundation.

\appendix \newpage

\section{Classical 1d harmonic oscillator} \label{app_classical1dHO}

In this appendix, we derive the evolution operators corresponding to the classical equations of motion of a one-dimensional harmonic oscillator, which is used in Eq.~\ref{eq_usingclassicalHO} of the main text to obtain the trajectories shown in Fig.~\ref{fig_classicalsolutions}b. We thus start from a Hamiltonian 
\begin{equation}
H = \frac{p^2}{2m} + \frac{1}{2} k r^2 , 
\end{equation}
where $m$ and $k$ are allowed an arbitrary sign, and with define the possibly complex eigenfrequency $\omega = \sqrt{k/m}$. In terms of the reduced variables 
\begin{equation}
\tilde{r} = \sqrt{m\omega} r , \quad \tilde{p} = p/\sqrt{m\omega} , 
\end{equation}
Hamilton's equations of motion read
\begin{equation}
\frac{{\rm d}}{{\rm d}t} \begin{bmatrix} \tilde r \\ \tilde p \end{bmatrix} = \omega  \begin{bmatrix} 0&1\\-1&0 \end{bmatrix} \begin{bmatrix} \tilde r \\ \tilde p \end{bmatrix} 
\end{equation}
and can be straightforwardly solved by exponentiation 
\begin{equation}
\begin{bmatrix} \tilde r (t) \\ \tilde p (t) \end{bmatrix} = \begin{bmatrix} \cos \omega t& \sin\omega t\\-\sin\omega t&\cos\omega t \end{bmatrix} \begin{bmatrix} \tilde r (0) \\ \tilde p (0) \end{bmatrix} ,
\end{equation}
which we stress is valid both for a real and purely imaginary $\omega$. Going back to the original variable leads to 
\begin{equation}
\begin{bmatrix}  r (t) \\  p (t) \end{bmatrix} = \begin{bmatrix} \cos \omega t&\frac{1}{m\omega} \sin\omega t\\- \frac{k}{\omega}\sin\omega t&\cos\omega t \end{bmatrix} \begin{bmatrix}  r (0) \\  p (0) \end{bmatrix} .
\end{equation}

\section{Evolution of phase space densities} \label{app_phasespacedensities}

In this appendix, we derive the parameters of the elliptic isodensity lines of the phase-space distributions studied in Sec.~\ref{ssec:phasespaceevolution}. We recall that, following the classical equation of motion, the phase-space distribution at time $t$ only depends on the following variable 
\begin{equation} 
E_{t} = \frac{1}{2} \begin{bmatrix} \vect{r} & \vect{p} \end{bmatrix} Q \begin{bmatrix} \vect{r} \\ \vect{p} \end{bmatrix} , \quad Q = \begin{bmatrix} Q_{rr} & Q_{rp} \\ Q_{pr} & Q_{pp} \end{bmatrix} ,
\end{equation}
with the explicit form of $Q$ inferred from Eq.~\ref{eq_rotationcanonicaltransfo} and Eq.~\ref{eq_fulltimeevolution}
\begin{equation} 
Q = R^T \begin{bmatrix} U_+^T U_+ & 0 \\ 0 & U_-^T U_- \end{bmatrix} R, \quad R = \begin{bmatrix} c&0&0&s \\ 0&-s&c&0 \\ 0&c&s&0 \\ -s&0&0&c \end{bmatrix} . \notag
\end{equation}
The shift of momenta $\tilde{\vect{p}} = \vect{p} + Q_{pp}^{-1} Q_{pr} \vect{r}$ in the first line of Eq.~\ref{eq_densityrealspacedistribution} leads to 
\begin{equation}
\rho_t(x,y) = \int {\rm d}^2 \tilde{\vect{p}} \, f_0 \left( \vect{r}^T A^{-1} \vect{r} + \tilde{\vect{p}}^T Q_{pp} \tilde{\vect{p}} \right) , 
\end{equation} 
with 
\begin{equation}
A^{-1} = Q_{rr} - Q_{rp} Q_{pp}^{-1} Q_{pr} = [(Q^{-1})_{rr}]^{-1} ,
\end{equation}
where the second equality can be either checked with lengthy by straightforward direct calculation, or derived using standard results on Schur complements. The matrix inverse $Q^{-1}$ can be independently obtained as
\begin{equation}
Q^{-1} = R^T \begin{bmatrix} (U_+^T U_+)^{-1} & 0 \\ 0 & (U_-^T U_-)^{-1} \end{bmatrix} R , 
\end{equation}
which can be efficiently evaluated using the relation 
\begin{equation}
(U_\pm^T U_\pm)^{-1}= - Y (U_\pm^T U_\pm) Y , \quad Y = \begin{bmatrix} 0&1\\-1&0 \end{bmatrix} . 
\end{equation}
Acting on the $Y$ matrices on the rotations matrices, we find 
\begin{equation}
(Q^{-1})_{rr} = Q_{pp} \quad \Longrightarrow \quad A = Q_{pp} . 
\end{equation}

As explained in the main text, all parameters of the ellipse characterizing the real-space density at time $t$ can therefore be obtained from the eigen-decomposition of $Q_{pp}$. We now find this decomposition and derive the expression given in Eq.~\ref{eq_classicaldistribellipseparams}. For simplicity, we decompose $Q_{pp} = a_\mu \sigma^\mu$ onto Pauli matrices $\sigma^{\mu=0,1,2,3}$, and find from the definition of $Q$ above the explicit form
\begin{align}
a_0 & = 1 - \frac{\varepsilon^2}{4 \Omega_R} (\tau_+^2 - \tau_-^2) , \\
a_1 & = - \frac{\varepsilon \Omega}{2 \Omega_R} (c_- \tau_- - c_+ \tau_+) , \\
a_2 & = 0 , \\
a_3 & = -\frac{\varepsilon}{2} \left[ \left( 1+\frac{\Omega^2}{\Omega_R} \right) \tau_+^2 + \left( 1-\frac{\Omega^2}{\Omega_R} \right) \tau_-^2 \right] ,
\end{align}
where we recall that $c_\pm =\cos (\omega_\pm t)$, $\tau_\pm = \sin (\omega_\pm t) / \omega_\pm$ and $\Omega_R^2 = \Omega^2 + (\varepsilon/2)^2$. The tilt of the ellipse, measured by the $\phi$ from the $y$-axis, and the minor and the principal axis lengths can be obtained from these coefficients as 
\begin{widetext}
\begin{eqnarray}
\tan ( 2 \phi ) &=& - \frac{a_1}{a_3} = \frac{\Omega}{\Omega_R} \frac{c_+ \tau_+ - c_- \tau_-}{\left( 1+\frac{\Omega^2}{\Omega_R} \right) \tau_+^2 + \left( 1-\frac{\Omega^2}{\Omega_R} \right) \tau_-^2} , \\
\lambda_\pm^2 &=& a_0 \pm \sqrt{a_1^2+a_3^2} = 1 - \frac{\varepsilon^2}{4 \Omega_R} (\tau_+^2 - \tau_-^2) \pm \frac{\varepsilon}{2} \sqrt{\left[ \left( 1+\frac{\Omega^2}{\Omega_R} \right) \tau_+^2 + \left( 1-\frac{\Omega^2}{\Omega_R} \right) \tau_-^2 \right]^2 + \frac{\Omega^2}{\Omega_R^2} (c_+ \tau_+ - c_- \tau_-)^2 } , \notag
\end{eqnarray}
\end{widetext}
leading to the results quoted in Eq.~\ref{eq_classicaldistribellipseparams} of the main text. There, for short-handedness, we gave the expression of $\lambda_\pm^2$ using the relation 
\begin{equation}
\lambda_\pm^2 = a_0 \pm |a_1|\sqrt{1+\frac{1}{\tan^2 (2\phi)}} = a_0 \pm \frac{|a_1|}{|\sin (2\phi) |} . 
\end{equation}

\section{Time evolved Fock states} \label{app_Fockstates}

In this appendix, we use the expression of $A_\pm(t)$ in terms of the complex coordinate $\xi, \bar\xi$ (Eqs.~\ref{eq_cyclotrontimeevolved} and~\ref{eq_guidingtimeevolved}) to derive the explicit real-space representation of the time evolved Fock states algebraically defined in Eq.~\ref{eq_algebraicdeffockstates}. 

For completeness and readability, we first provide a more detailed description of the new holomorphic system of coordinates introduced in Eq.~\ref{eq_newholomorphic}. We have first used the complex representation
\begin{equation*}
	\begin{cases} z & = x+iy \\ \bar{z} & = x-iy \\  x & = \frac{1}{2}(z+\bar{z}) \\ y & = \frac{1}{2i}(z-\bar{z}) 
	\end{cases} \, , \quad 
	\begin{cases} p_z & = \frac{1}{2} (p_x - ip_y) \\ p_{\bar{z}} & = \frac{1}{2} (p_x + ip_y) \\  p_x & = p_z + p_{\bar{z}}  \\  p_y & = i(p_z - p_{\bar{z}})
	\end{cases} \, ,
\end{equation*}
well-suited to describe right and left moving particles in a magnetic field. Then, we obtain elliptic cyclotron orbits which motivates the definitions of stretched complex coordinates
\begin{equation*}
\begin{cases} \xi & = \alpha(f_+ z +g_+ \bar{z}) \\ \bar{\xi} & = \alpha(g_+^* z + f_+^* \bar{z} )
\end{cases} \, , \quad 
\begin{cases} \alpha p_\xi & = f_+^* p_z - g_+^* p_{\bar{z}} \\ \alpha p_{\bar{\xi}} & = f_+ p_{\bar{z}} - g_+ p_z 
\end{cases} \, .
\end{equation*}
made to enable the simple expression $A = \frac{1}{2}(\xi + 2i p_{\bar{\xi}})$ and to best match the geometry of the cyclotron orbit at any point in time.

To get the explicit form of Fock states, we first recall the formula obtained for the vacuum state 
\begin{equation}
\phi_t(\xi, \bar\xi) = \braket{x,y}{0,0}_t = \frac{1}{\sqrt{\pi|u|}} \exp\left[ \frac{\delta \xi^2 - |\xi|^2}{2} \right] . 
\end{equation}
Using the identity 
\begin{equation}
A_-(t) \phi_t(\xi, \bar\xi) = \phi_t(\xi, \bar\xi)  \left( \frac{\xi}{u} - u \partial_{\bar{\xi}} - v \partial_\xi \right) ,
\end{equation}
multiple times offers the following expression for the guiding center Fock states in the lowest Landau level
\begin{equation}
\braket{x,y}{0,n_-}_t = \frac{\phi_t(\xi,\bar\xi)}{\sqrt{n_-!}} \left( \frac{\xi}{u} - v\partial_\xi  \right)^m \underline{1} \, ,
\end{equation}
with $\underline{1}$ the function everywhere equal to one. We recognize the definition of Hermite's polynomials $\{H_n\}_n$ and conclude
\begin{equation} \label{appeq_AnisotropicLowestLandauLevel}
\braket{x,y}{0,n_-}_t = \frac{\phi_t(\xi,\bar\xi)}{\sqrt{n_-!}} \left( \frac{\delta}{2} \right)^{\frac{n_-}{2}} H_{n_-} \left( \frac{\xi}{\sqrt{2uv}} \right) \, .
\end{equation}
Similarly, we can combine the identities 
\begin{equation}
A_+^\dagger \phi_t(\xi,\bar\xi)  = \phi_t(\xi,\bar\xi) [(\bar{\xi}-\delta\xi)-\partial_\xi] , 
\end{equation}
\begin{equation*}
[(\bar{\xi}-\delta\xi)-\partial_\xi] H_k(a\xi) = H_k(a\xi) [(\bar{\xi}-\delta\xi)-\partial_\xi]  - 2ak H_{k-1}(a\xi)
\end{equation*}
to find, after careful calculations, the expression
\begin{widetext}
\begin{equation}\label{appeq_RealSpaceExpression_GenericFockState}
\braket{x,y}{n_+,n_-}_t  = \frac{i^{n_+} \phi_t(\xi,\bar\xi)}{\sqrt{n_+!n_-!}} \left(\frac{\delta}{2}\right)^{\frac{n_-+n_+}{2}} \sum_{k=0}^{{\rm min}(n_+,n_-)} k! \binom{n_+}{k} \binom{n_-}{k} \left(\frac{2i}{v}\right)^k H_{n_--k}\left(\frac{\xi}{\sqrt{2uv}}\right) H_{n_+-k}\left(\frac{\bar{\xi}-\delta\xi}{i \sqrt{2\delta}}\right) \, ,
\end{equation}
\end{widetext}
which generalizes the $n_+=0$ solution of Refs.~\cite{Fetter_ClassicalOrbitsAndLLL,Oktel_WFandVortices}.

\bibliography{biblio}

\end{document}